\documentclass[aps,prl,toolkits,twocolumn]{revtex4}

\usepackage{mathrsfs}
\usepackage{amsmath}

\newcommand\Dp{{\Delta_+}}
\newcommand\Dm{{\Delta_-}}
\newcommand\hfrac[2]{{#1}/{#2}}
\renewcommand\section[1]{{\em #1 ---}}
\newcommand\paper{paper}

\begin{document}

\title{Holography and the speed of sound at high temperatures}

\author{Paul M. Hohler}
\email{pmhohler@uic.edu}

\affiliation{Department of Physics, University of Illinois,
Chicago, IL 60607-7059, USA}

\author{Mikhail A. Stephanov}
\email{misha@uic.edu}

\affiliation{Department of Physics, University of Illinois,
Chicago, IL 60607-7059, USA}

\begin{abstract}
  We show that in a
  general class of strongly interacting theories at high
  temperatures the speed of sound approaches the conformal
  value $c_s^2=1/3$ universally from {\em below}.
  This class includes theories
  holographically dual to a theory of gravity coupled to a single
  scalar field, representing the operator of the scale anomaly.
\end{abstract}


\maketitle

\section{Introduction}
\label{sec:intro}
The discovery of the holographic correspondence between string and
gauge theories \cite{Maldacena:1997re,Gubser:1998bc,Witten:1998qj} has
led to a number of applications to modelling properties of quark-gluon
plasma of QCD in the regime relevant to heavy-ion collision
experiments (see, e.g., Ref.~\cite{Gubser:2009md} for a guide
and references).

Although the precise form of the theory holographically dual to QCD is
not known, many features of QCD can be implemented on the string
theory side of the correspondence. Alternatively, one can pursue a
bottom-up approach, by constructing an effective theory, or a model,
with a {\em minimal} set of operators needed to describe the relevant
physics. In this {\paper}, we consider a general class of such models,
describing the physics of scaling violation in QCD.
In
Refs.~\cite{Gubser:2008ny,Gubser:2008yx,Gursoy:2008za} such theories
have been considered with the aim to model, or mimic, the equation of
state of QCD known numerically from the lattice calculations. This has
been achieved by tuning the potential $V(\phi)$ of the scalar field in
the model.

The main goal of this {\paper} is to establish properties of the
equation of state $p(\epsilon)$, which are {\em universal} with respect
to the choice of the potential $V(\phi)$. An obvious universal
property is that the conformal symmetry is gradually restored, and
thus $p(\epsilon)\to\epsilon/3$, as the temperature $T\to\infty$.
This is natural since the temperature becomes the only relevant
scale. It is a welcomed feature of the models since the same
phenomenon occurs in QCD.

Here,
we shall show that the {\em deviation} of the speed of sound
$c_s^2=dp/d\epsilon$ from the conformal limit $c_s^2=1/3$ is
universally {\em negative} in such holographic models, at least to the
leading order in inverse temperature.
  
At the outset, we must point out that $c_s^2=1/3$ is
by no means a universal upper bound on the sound velocity. A
counterexample is given by the speed of sound in QCD at large isospin
chemical potential \cite{Son:2000by}, where $c_s$ can approach the
speed of light in a certain limit. Nevertheless, a number of string
theory examples~\cite{Buchel:2003ah,Buchel:2004hw,Benincasa:2005iv} of
holographically dual theories do indeed consistently show $c_s^2\le1/3$.

In QCD, Monte Carlo lattice
calculations ~\cite{Karsch:2007dp} show that this inequality is
fulfilled. In the regime $T\gg\Lambda_{\rm QCD}$, this
inequality follows from asymptotic freedom: $\beta(\alpha)<0$. For a pure $N_c$-color Yang-Mills
theory at high $T$, for example, \cite{Kapusta-book}
\begin{equation}
  \label{eq:cs-beta}
    c_s^2\simeq 1/3+5 N_c/(36\pi)\,\beta(\alpha)<1/3.
\end{equation}

How general is the inequality  $c_s^2<1/3$ and what are the
prerequisites for it to hold? For example, does it hold in
regimes where the theory is close to being conformal, but is still
strongly interacting, as is apparently the case for QCD in the range of $T\sim
(2-3)\,T_c$?

To investigate this question, we consider a class of theories, or models,
which possess a holographically dual description. The minimal set of
operators that we need to consider consists of the stress-energy
tensor $T^{\mu\nu}$ and a scalar operator ${\cal O}$ necessary
to provide a nonzero right-hand side of the scale (trace) anomaly
equation. As an example, in QCD, the role of such an operator is
played by ${\rm Tr} F^2$: $\theta=\beta(\alpha)/(8\pi\alpha^2)\,{\rm
  Tr}F^2$.

At this point it is helpful to express the energy density
$\epsilon$ and pressure $p$ in terms of the heat function (enthalpy)
$w=\epsilon+p$ and the trace of the stress-energy tensor
$\theta=\epsilon-3p$:
\begin{equation}
  \label{eq:ep-wt}
  \epsilon=(3w+\theta)/4;\qquad
  p = (w-\theta)/4.
\end{equation}
Then
\begin{equation}
  \label{eq:cs-theta}
  c_s^2=\frac{dp}{d\epsilon}=\frac{1-d\theta/dw}{3+d\theta/dw}\,.
\end{equation}
In conformal theories $\theta\equiv0$ and $c_s^2=1/3$. The inequality
$c_s^2<1/3$ is equivalent to $d\theta/dw>0$.
This inequality is somewhat reminiscent, but not equivalent, to the inequality
$d(p\,T^{-4})/dT=\theta/T^5>0$ conjectured in
Ref.\cite{Bjorken:1982qr} and the weaker constraint $\int\! dT\,\theta/T^5
>0$ proposed in
Refs.~\cite{Appelquist:1999hr,Appelquist:1999vs}.

Since the enthalpy $w$ appears to play here the role of a more natural
thermal parameter than $T$, it is helpful to keep in mind that $w$ is a monotonous
function of $T$: $dw/dT=c_v+s>0$.

\section{The theory}
\label{sec:theory}
The class of the five-dimensional (5D) holographic theories that we consider
contains a scalar field $\phi$, holographically dual to the (scaling
violating) operator ${\cal O}$, coupled to the metric $g_{\mu\nu}$ (dual
to $T^{\mu\nu}$). The fields are governed by the action
\begin{equation}\label{eq:S5}
\begin{split}
S_5 = \frac{1}{2 \kappa^2} \biggl(&\int_M \!\! d^5x \; \sqrt{-g}\, \Bigl (R - V(\phi) - \frac{1}{2} (\partial \phi)^2 \Bigr)\\& - 2 \int_{\partial M} \!\!\!\! d^4x \; \sqrt{-\gamma}\, K \biggr),
\end{split}
\end{equation}
where $R$ is the Ricci scalar, $g$ is the determinant of the metric,
$\gamma$ is the determinant of the induced metric on the UV boundary
$\partial M$, $K$ is the extrinsic curvature on $\partial M$, and
$\kappa^2$ is the 5D Einstein gravitational constant. The value of
$\kappa^2$ is inversely proportional to the number of the degrees of
freedom in the dual four-dimensional theory, e.g., $N_c^2$ in a gauge theory with
large number of colors $N_c$. The smallness of $\kappa^2$ (i.e.,
the largeness of the number of colors) controls the semiclassical
approximation which we use. The last term in the action is the
Gibbons-Hawking term, which removes the boundary terms arising upon
integration by parts of the terms in $R$ linear in second derivatives
of the metric~\cite{Gibbons:1976ue}. This boundary term does not
affect classical equations of motion, but is essential for evaluating
variations of the action with respect to the boundary values of the
fields.

We assume that the potential $V(\phi)$ has an extremum at
$\phi=0$. The value of $V(0)=2\Lambda=-12$, where $\Lambda$ is the
cosmological constant necessary to achieve proper asymptotics of the
metric near the boundary.

It should be noted that we are going to treat the holographic theory
with a method technically different from the one used in either
Refs.~\cite{Gubser:2008ny,Gubser:2008yx} or
Ref.~\cite{Gursoy:2008za}. Our approach is complementary to the existing
ones and has its own advantages, which might go beyond the specific
application which we consider in this {\paper}.

According to the holographic correspondence, the correlation functions of
the dual four-dimensional (4D) theory are equal to the variations of the 5D action under
the changes of the boundary conditions on the 5D fields. We shall first
determine the extremum of the action, and then consider first-order
variations, which we can relate to energy density and pressure. The
most general metric (up to general coordinate transformations) 
possessing three-dimensional (3D) Euclidean isometry is given by
\begin{equation} \label{eq:length}
ds^2 = \frac{1}{z^2}\left(-f(z) dt^2 + d\vec{x}^2\right) + e^{2 B(z)} \frac{dz^2}{z^2 f(z)}.
\end{equation}
The functions $B(z)$ and $f(z)$ depend on the holographic
coordinate $z$ and will be determined by extremizing the action.
The boundary $\partial M$ is located at $z = \varepsilon$ with $\varepsilon$ acting
as an UV regulator. In relations involving only physical quantities $\varepsilon$ can be
taken to $0$.

Substituting the metric of Eq.~(\ref{eq:length}) into the equations of
motion we find:
\begin{eqnarray}
&\dot{B} = -\frac{1}{6}\, \dot{\phi}^2, \label{eq:ein1}\\
&\ddot{f} = \left({4} + \dot{B}\right) \dot{f}, \label{eq:ein2}\\
&-6 \dot{f} + f \left({24} - \dot{\phi}^2\right) + 2 {e^{2 B}}\, V(\phi) = 0, \label{eq:ein3}\\
&\ddot{\phi}f + \dot{\phi}\left(\dot{f} - f(4 + \dot{B})\right) - e^{2 B}\, V^\prime(\phi) = 0, \label{eq:ein4}
\end{eqnarray}
where a dot denotes a $\log z$ derivative, e.g.,
$\dot\phi=z\,d\phi/dz$, while $V'=dV/d\phi$. 
The holographic correspondence provides the boundary conditions at
the UV boundary $z=\varepsilon$. The boundary condition on $\phi$,
\begin{equation}
  \label{eq:phi-bc}
  \phi(\varepsilon)=c\,\varepsilon^{\Delta_-}\,,
\end{equation}
corresponds to introducing a term $c\,{\cal O}$ into the action of the
4D theory dual to the theory in Eq.~(\ref{eq:S5}),
where $c$ is the source of the scalar operator ${\cal O}$, dual to the
field $\phi$. The dimensions of the source and the operator are given
by
\begin{equation}
  \label{eq:cOdelta}
  [c]=\Dm;\quad [{\cal O}]=\Dp;\quad \Dp+\Dm=4.
\end{equation}

The metric must approach the
Lorentz invariant form at the UV boundary, hence,
\begin{equation}
  \label{eq:f-bc}
  f(\varepsilon)=1\,.
\end{equation}
Equation (\ref{eq:ein2}) can be integrated once to give
\begin{equation} \label{eq:ein2b}
\dot{f} = - w z^4 e^{B}.
\end{equation}
The integration constant $w$ must be positive if the metric is to
possess a horizon $f(z_H)=0$ at some value of $z_H$.

There are no more independent boundary conditions. The boundary value of $B$ is
determined by Eq.~(\ref{eq:ein3}), which is algebraic in $B$.
The role of the
second boundary condition for
Eq.~(\ref{eq:ein4}) is played by the requirement
that $\phi$ is finite at the horizon, $z=z_H$, which is a
regular singular point of the second order differential Eq.~(\ref{eq:ein4}).

Thus, we find a two-parameter family of solutions. These parameters are
$c$ from Eq.~(\ref{eq:phi-bc}) and $w$ from
Eq.~(\ref{eq:ein2b}). Different members of this family are related by
rescaling $z\to z/\lambda$, $w\to\lambda^{4}w$, $c\to
\lambda^{\Delta_-}c$, which represents scale invariance inherent in the
action given by Eq.~(\ref{eq:S5}).

Near the $z=\varepsilon\to0$ boundary, $\phi\to0$, $B\to0$, $\dot B\to
0$ and $\dot f\to 0$.  Equation~(\ref{eq:ein4}) for $\phi$ can be
linearized and the asymptotic behavior of $\phi$ near the boundary can
be determined easily:
\begin{equation} \label{eq:asy}
\phi(z) \to (c - d\, \varepsilon^{\Delta_+-\Delta_-})\,z^{\Delta_-} 
+ d\, z^{\Delta_+} + \ldots,
\end{equation}
where the curvature of the potential $V''(0)\equiv m^2$ determines the
indices $\Delta_\pm=2\pm\sqrt{4+m^2}$. The coefficient of the first
term is related to $c$ by Eq.~(\ref{eq:phi-bc}). The coefficient $d$ of the
second linearly independent solution should be determined by the finiteness
condition at the horizon.

By calculating the derivative of the 5D action with respect to $c$ and
matching it, by holographic correspondence, to the expectation value
$\langle{\cal O}\rangle$, one finds \cite{Klebanov:1999tb}
\begin{equation} \label{eq:op}
\langle \mathcal{O} \rangle = -\hfrac{\partial S_5}{\partial c} 
= -d \, (\Delta_+-\Delta_-).
\end{equation}

\section{Equation of state}
\label{sec:eos}
Rather than following existing approaches to calculating the speed of
sound, based on the relation $c_s^2=d \log T/d \log s$,
\cite{Gubser:2008ny,Gubser:2008yx} or looking at the poles of the
two-point function for the sound channel
\cite{Policastro:2002tn,Kovtun:2005ev}, we shall directly calculate
{\em one}-point functions $\langle T^{tt}\rangle$ and $\langle
T^{xx}\rangle$ and use the relation $c_s^2=dp/d\epsilon$. As we shall
see, this quickly yields some closed-form results, not evident in the
alternative approaches.

In order to calculate the one-point function, such as $\langle
T^{tt}\rangle$, we observe that the generalization of the 4D theory to a
curved 4D background with metric $h_{\mu\nu}$ corresponds,
holographically, to imposing the following boundary condition on the
bulk 5D metric% (compare to Eq.~(\ref{eq:phi-bc})):
\begin{equation}
  \label{eq:gmunu-bc}
  g_{\mu\nu}(\varepsilon)=h_{\mu\nu}\,\varepsilon^{-2}\,.
\end{equation}
Holographic correspondence then gives us
\begin{equation}
  \label{eq:ds-dh}
   \langle
T^{\mu\nu}\rangle=2\,\delta S_5/\delta h_{\mu\nu}\,.
\end{equation}
Variation of the boundary
condition causes variations of the metric in the bulk, as it follows
the equations of motion. But since we are varying
around the extremum of the action, the only contribution to the first-order
variation comes from the boundary. This contribution consists of the
variation of the boundary terms, which appear during integration by
parts while deriving equations of motion for, e.g., $g_{tt}$, as well as the
variation of the Gibbons-Hawking term. Altogether this gives
\begin{equation}\label{eq:TttTxx}
\langle T^{tt} \rangle = -\frac{6}{\varepsilon^4}
  e^{-B(\varepsilon)}; 
 \qquad \langle T^{xx} \rangle = w -  \langle T^{tt} \rangle .
\end{equation}
Both of these quantities are divergent as $\varepsilon \rightarrow
0$. This represents the familiar vacuum energy divergence in the
quantum field theory. A simple vacuum subtraction
\begin{equation}
\begin{split}
\epsilon &= \langle T^{tt} \rangle - \langle T^{tt} \rangle_{T=0}\,,\\
p &= \langle T^{xx} \rangle - \langle T^{xx} \rangle_{T=0},
\end{split}
\end{equation}
takes care of this, and we find for the energy density and the
pressure, after solving Eq.~(\ref{eq:ein3}) for $B$ at
$z=\varepsilon\to0$ with $\phi$ from Eq.~(\ref{eq:asy}):
\begin{equation} \label{eq:ep}
\begin{split}
\epsilon &= (\,3w - c\, d \,\Delta_-\, (\Delta_+-\Delta_-)\,)/4\,,\\
p &= (\,w +  c\, d \,\Delta_-\, (\Delta_+-\Delta_-)\,)/4.
\end{split}
\end{equation}
We can now conclude that the integration constant introduced in
Eq.~(\ref{eq:ein2b}) is the enthalpy, $w=\epsilon+p$, of the
corresponding 4D field theory. We also observe that the scale anomaly
\begin{equation}
  \label{eq:theta-cd}
  \theta = \epsilon-3p = - c\, d \,\Delta_-\, (\Delta_+-\Delta_-)
=  \Delta_-\, c\,\langle{\cal O}\rangle,
\end{equation}
is related to the expectation value of the operator ${\cal O}$ [using
Eq.~(\ref{eq:op})]. This relation is easy to derive directly on the
field theory side by observing that the violation of the scale
invariance comes from the dimensionful parameter $c$.  Naturally, it
is proportional to the dimension of $c$, $\Delta_-$.

From Eq.~(\ref{eq:ep}), the speed of sound can be expressed now in
terms of the derivative $\partial d/\partial w$ at fixed $c$ and
dimensions $\Delta_\pm$:
\begin{equation} \label{eq:cs2}
\begin{split}
c_s^2 = \frac{d p}{d \epsilon} & = 
\frac{1+c \,\Delta_-\, (\Delta_+-\Delta_-)\,(\partial d/\partial w)}
{3- \,c\,\Delta_-\, (\Delta_+-\Delta_-)\,(\partial d/\partial w) }\,.\\ 
% &= \frac{1}{3} \frac{(1+ c \,(4 - \Delta) \frac{\delta \langle \mathcal{O} \rangle}{\delta w})}{(1-\frac{1}{3} \,c\, (4 - \Delta)  \frac{\delta \langle \mathcal{O} \rangle}{\delta w})},
\end{split}
\end{equation}
It should be emphasized that this equation is an {\em exact}
expression valid for {\em all} temperatures.

\section{High temperature}
\label{sec:highT}
The high-temperature limit is equivalent to the small $c$ limit, since $c$ is
the only other dimensionful parameter in the theory, and high-temperature expansion is controlled by the dimensionless parameter
$c/T^{\Delta_-}$, or $c\,w^{-\Dm/4}$. In this limit,
\begin{equation}
\begin{split}
c_s^2 &= 1/3  + ({4}/{9})\, c\,\Delta_-\, (\Delta_+-\Delta_-)\,(\partial d/\partial w) + \ldots. \\
\end{split}
\end{equation}
We shall now show that, even though generally the dependence of $d$
on $w$ can be only found numerically, in the high-temperature (i.e., high
$w$) limit an analytic expression can be found for arbitrary potential
$V(\phi)$.

One can begin by observing that at large $w$ the function $f$ varies
very rapidly according to Eq.~(\ref{eq:ein2b}). This means one can
neglect variations of the function $B$ between the boundary $z=\varepsilon$
and the horizon $z=z_H$, since $z_H$ becomes small (as $ w^{-1/4}$). 
Since on the boundary $B=0$ (up to 
terms of order $\varepsilon^{2\Delta_-}$, negligible here, according to
Eq.~(\ref{eq:ein3})), we find from Eq.~(\ref{eq:ein2b})
\begin{equation}
  \label{eq:fz}
   f(z) = 1 - w\, z^4/4.
\end{equation}
Another consequence is that $\phi$, which is small at $z=\varepsilon$,
remains small up to $z_H$ ($\phi\sim cz_H^{\Delta_-}\sim
c/T^{\Delta_-}\ll1$), and the linearized approximation to
Eq.~(\ref{eq:ein4}) is valid not only near the boundary, but all the way to the
horizon. With $B=0$ and $f$ from Eq.~(\ref{eq:fz}) we obtain
\begin{equation} \label{eq:dphi}
\left(1 - \frac{1}{4}\, w z^4\right) {\phi}''
- \left (  \frac{3}{z} + \frac{w z^4}{4} \right) {\phi}' 
- \frac{m^2}{z^2}\,\phi = 0.
\end{equation}
Equation (\ref{eq:dphi}) can be solved analytically
\begin{equation} \label{eq:phi}
\begin{split}
\phi(z) & = c\,z^{\Dm}\; {}_2F_1\left(\hfrac{\Dm}{4},\,\hfrac{\Dm}{4},\,\hfrac{\Dm}{2},\,w z^4\right/4)\\
& + d \, z^{\Dp} \;{}_2F_1\left(\hfrac{\Dp}{4},\,\hfrac{\Dp}{4},\,\hfrac{\Dp}{2},\, w z^4/4\right),
\end{split}
\end{equation}
where the coefficients follow the notations of Eq.~(\ref{eq:asy}) (up
to terms ${\cal
  O}(\varepsilon^{\Delta_+-\Delta_-})$, here negligible). Both linearly independent solutions are
logarithmically divergent at the horizon $z=z_H$, where
$wz_H^4/4=1$. The condition $|\phi(z_H)|<\infty$ requires us to select
the linear combination in which these divergences cancel. This fixes
$d$ in terms of $c$:
\begin{equation} \label{eq:dc}
\begin{split}
d &= - c\; w^{{(\Dp-\Dm)}/{4}}\, D(\Dm)\,,\\ 
\end{split}
\end{equation}
where the function $D(\Dm)=1/D(\Dp)$ is given by
\begin{equation} \label{eq:DD}
\begin{split}
D(\Dm)
% &= \frac{2-\Dp}{\pi \,2^{\Dp}} \, 
%\tan\left(\pi \Dp/4\right) 
%\frac{\Gamma(\Dp/4)^4}{\Gamma(\Dp/2)^2}\\
&=\frac{\pi \,2^{\Dm}}{2-\Dm}
\cot\left(\pi \Dm/4\right)
\frac{\Gamma(\Dm/2)^2}{\Gamma(\Dm/4)^4}
%= \frac1{D(\Dp)}
\,.
\end{split}
\end{equation}
Substituting Eq.~(\ref{eq:dc}) into Eq.~(\ref{eq:cs2}) we find our main result
\begin{equation} \label{eq:csW}
c_s^2 = \frac{1}{3} - \frac{1}{9}\, c^2 \, \Dm (\Dp-\Dm)^2  w^{-\Dm/2} D(\Dm) + \ldots.
\end{equation}
It is clear that the correction term is {\em negative} for all values of
$0<\Dm<2$ (i.e., $2<\Dp<4$). As expected, the correction vanishes with the power
of $w$ dictated by the dimension of the operator ${\cal O}$.

\section{Discussion}
\label{sec:conclusion}
When this work was completed, the authors learned about a similar
result \cite{Cherman:2009}, obtained using the relation
 $c_s^2=d\log T/d\log s$. In Ref.~\cite{Cherman:2009} the speed of
sound is expressed in terms of the value $\phi_H$ of the scalar field
at the horizon (since both $s$ and $T$ are
calculated at the horizon): $c_s^2=1/3-C\phi_H^2+\ldots\,$. The coefficient
$C$ is then given by an integral from the boundary to the horizon of
the square of a hypergeometric function, which is manifestly
positive. We have verified that, upon evaluating the integral, the
result of Ref.~\cite{Cherman:2009} coincides with Eq.~(\ref{eq:csW}). For
completeness of comparison we evaluate $\phi_H$ explicitly by using
Eq.~(\ref{eq:phi}) and applying relation Eq.~(\ref{eq:dc})
\begin{equation}\label{eq:phi-h}
\phi_H = c \;w^{\hfrac{\Delta}{4}-1}\; 2^{-\hfrac{\Delta}{2}} 
(2\Delta-4)\,
 \frac{\Gamma(\hfrac{\Delta}{4})^2}{\Gamma(\hfrac{\Delta}{2})}.
\end{equation}
where $\Delta=\Dp$. Using Eq.~(\ref{eq:csW}) we then find
\begin{equation}\label{eq:csphi}
c_s^2 = \frac{1}{3} - \frac{1}{18 \pi}\,(4-\Delta) \,(4-2 \Delta)\, \tan\left(\hfrac{\pi \Delta}{4} \right)\, \phi_H ^2 + \ldots\,.
\end{equation}

As we have pointed out already, and as Eq.~(\ref{eq:phi-h})
demonstrates explicitly, in the high-temperature (large~$w$) limit the
value of $\phi$ remains small everywhere between the boundary and the
horizon. This is the origin of the universality we find. The speed of
sound near the high $T$ limit depends only on the behavior of the scalar
potential near the origin, i.e., specifically, on
$V''(0)=m^2=\Delta(\Delta-4)$.

One can interpret and further generalize our results in the following
way. In the high-temperature limit, Eq.~(\ref{eq:theta-cd}) becomes
$\theta=\Dm\,c^2\,\chi_{{\cal O}}+\ldots\ $. The susceptibility
$\chi_{{\cal O}}\equiv\partial\langle{\cal O}\rangle/\partial c$ can
be related to the effective potential for $\langle{\cal O}\rangle$,
defined as the Legendre transform of the generating functional $W(c)$
(in holography $W=-S_5$), i.e.,
$\Gamma(\langle{\cal O}\rangle)\equiv W(c)+c\langle{\cal O}\rangle$,
as $\chi_{{\cal O}}=1/\Gamma''(0)$.  Stability implies
$\Gamma''(0)>0$, and thus $\chi_{{\cal O}}>0$. 
Consequently $\theta>0$, as conjectured in~\cite{Bjorken:1982qr}.
Holographic models confirm these expectations according to
Eqs.~(\ref{eq:op}) and~(\ref{eq:dc}).
% $\chi_{{\cal O}}=(\Dp-\Dm)\,w^{(\Dp-\Dm)/4}\,D(\Delta_-)>0$. 

We also find that $d\theta/dw>0$, which requires $d\chi_{{\cal
    O}}/dw>0$. That means that the curvature $\Gamma''(0)=\chi_{{\cal
    O}}^{-1}$ {\em decreases} with temperature. This behavior is
unusual, if one recalls that in a weakly coupled $\lambda\phi^4$
scalar theory, the leading perturbative temperature correction to the
curvature ($\sim\lambda T^2$) increases with temperature (e.g.,
the restoration of a broken symmetry is a well-known
manifestation of this).  The opposite behavior of the curvature of
the effective potential $\Gamma(\langle{\cal O}\rangle)$ can be
understood by counting dimensions:  $[\chi_{{\cal O}}^{-1}]=\Dm-\Dp<0$, which
means $\chi_{{\cal O}}^{-1}\sim T^{\Dm-\Dp}$ and decreases with $T$.

In conclusion, we have shown that in a quite general class of gravity
dual theories with a single scalar operator representing the scale
anomaly the speed of sound always approaches the conformal value $c_s^2=1/3$
from {\em below}.

{\it Acknowledgments.}  We would like to thank A.~Cherman, T.~Cohen, and A.~Nellore for discussions and for sharing the
results of Ref.~\cite{Cherman:2009} prior to publication. The hospitality
of the Institute of the Nuclear Theory at the University of Washington
and partial support of the Department of Energy through the program
``Strings and Things'' during the early
stages of this work is gratefully acknowledged. This work is
supported by the DOE grant No.\ DE-FG0201ER41195.

\end{document}